\documentclass[lettersize,journal]{IEEEtran} 
\usepackage{cite}
\usepackage{mathrsfs}
\usepackage{amsthm}
\usepackage{etoolbox} \makeatletter \patchcmd{\@makecaption} {\scshape} {} {} {} \makeatother
\usepackage[cmex10]{amsmath}
\usepackage{mathtools}
 \usepackage{siunitx}
\usepackage{array}
\usepackage{color}
\usepackage{tabularray}
\usepackage{eqparbox}
\usepackage{bm}

\usepackage[table,dvipsnames]{xcolor}
\usepackage{multicol,booktabs,tabularx}

\usepackage{cases}
\usepackage{algorithm}
\usepackage{paralist}
\usepackage{algpseudocode}

\usepackage{graphicx}
\usepackage{subfigure}
\usepackage{epstopdf}
\usepackage{epsfig}
\usepackage{amssymb}
\usepackage{array}
\usepackage{multirow}

\title{Enhanced Fingerprint-based Positioning With Practical Imperfections: Deep learning-based approaches}
\author{Shugong Xu, \textit{Fellow, IEEE}, Jun Jiang, Wenjun Yu, Yilin Gao, Guangjin Pan, Shiyi Mu, Zhiqi Ai, Yuan Gao, Peigang Jiang, Cheng-Xiang Wang, \textit{Fellow, IEEE}

\thanks{This work was in part supported by Shanghai Natural Science Foundation under Grant 22ZR1422200, and in part by the 6G Science and Technology Innovation and Future Industry Cultivation Special Project of Shanghai Municipal Science and Technology Commission under Grant 24DP1501001. (Yuan Gao is the corresponding author)}

\thanks{Shugong Xu is with Xi’an Jiaotong-Liverpool University, Suzhou, China, email: shugong.xu@xjtlu.edu.cn.}

\thanks{Jun Jiang, Wenjun Yu, Yilin Gao, Shiyi Mu, Zhiqi Ai and Yuan Gao are with the School of Communication and Information Engineering, Shanghai University, China, email: jun\_jiang@shu.edu.cn, yuwenjun@shu.edu.cn, gaoyilin@shu.edu.cn, shiyimu@shu.edu.cn, aizhiqi-work@shu.edu.cn and gaoyuansie@shu.edu.cn.}

\thanks{Guangjin Pan is with the Department of Electrical Engineering, Chalmers University of Technology, Gothenburg, Sweden, email: guangjin.pan@chalmers.se. }

\thanks{Peigang Jiang is with Shanghai Research Institute, Huawei Technologies Company Limited, China email: jiangpeigang@huawei.com. }

\thanks{Cheng-Xiang Wang is with National Mobile Communications Research Laboratory, School of Information Science and Engineering, Southeast University, Nanjing, China, e-mail: chxwang@seu.edu.cn}
}

\begin{document}
\maketitle
\begin{abstract}
High-precision positioning is vital for cellular networks to support innovative applications such as extended reality, unmanned aerial vehicles (UAVs), and industrial Internet of Things (IoT) systems. Existing positioning algorithms using deep learning techniques require vast amounts of labeled data, which are difficult to obtain in real-world cellular environments, and these models often struggle to generalize effectively. To advance cellular positioning techniques, the 2024 Wireless Communication Algorithm Elite Competition\footnote{This competition was sponsored by Huawei Ltd. and further details can be found at https://wireless.yuny.top/\#/} was conducted, which provided a dataset from a three-sector outdoor cellular system, incorporating practical challenges such as limited labeled-dataset, dynamic wireless environments within the target and unevenly-spaced anchors, Our team developed three innovative positioning frameworks that swept the top three awards of this competition, namely the semi-supervised framework with consistency, ensemble learning-based algorithm and decoupled mapping heads-based algorithm. Specifically, the semi-supervised framework with consistency effectively generates high-quality pseudo-labels, enlarging the labeled-dataset for model training. The ensemble learning-based algorithm amalgamates the positioning coordinates from models trained under different strategies, effectively combating the dynamic positioning environments. The decoupled mapping heads-based algorithm utilized sector rotation scheme to resolve the uneven-spaced anchor issue. Simulation results demonstrate the superior performance of our proposed positioning algorithms compared to existing benchmarks in terms of the \{90\%, 80\%, 67\%, 50\%\} percentile and mean distance error.

\end{abstract}
\section{Introduction}

Positioning has been a crucial function of cellular networks since the introduction of 2G technology. As we look toward the future of Beyond 5G (B5G) and 6G, there is a strong emphasis on supporting emerging applications such as extended reality, unmanned aerial vehicles (UAVs), and smart environments, all of which demand sub-centimeter positioning accuracy\cite{gao2024performance,wang2023road}. Specifically, it is anticipated that horizontal positioning accuracy will reach between \SI{0.2}{\metre} and \SI{0.3}{\metre} in both outdoor scenarios (with mobility of up to \SI{250}{\kilo\metre/\hour}) and indoor environments (with mobility of up to \SI{30}{\kilo\metre/\hour})\cite{22_261}.

Fingerprint positioning is a notable method that involves measuring certain features of the wireless signal at known locations, referred to as reference points, during an offline phase. These measurements are compiled into a database known as the “radio map”. The user’s location is then estimated by comparing the measured signal features to the information stored in the radio map\cite{li2023automatic}. Initially, fingerprint generation relied on low-dimensional signal features, such as received signal strength (RSS)\cite{chen2019learning}, which often led to inadequate positioning accuracy. 

With the rapid advancement of artificial intelligence, high-dimensional signal features could be exploited\cite{he2016deep}, such as the angle and delay information of multi-path\cite{li2023automatic}. Deep-learning-based schemes, including convolutional neural network (CNN)-based scheme\cite{li2023automatic}, Transformer-based scheme and generative adversarial network (GAN)-based scheme\cite{hu20233d}, uncover the underlying relationships between physical locations and channel state information (CSI), facilitating more accurate positioning, at the expense of the following three primary limitations: 
\begin{inparaenum}[\itshape a\upshape)]
\item a significant dependence on large labeled CSI datasets, which are challenging to obtain in practical applications, especially with limited reference points;
\item uneven spatial distribution of reference points (RPs); and
\item inadequate generalization and robustness of models when dealing with positioning in the dynamic environment.
\end{inparaenum}

Recently, channel charting, a self-supervised learning technique, has emerged as a valuable tool for positioning. Channel charting maps high-dimensional CSI to low-dimensional virtual coordinates, ensuring that the dissimilarity of CSIs is a measure of the distance between two points\cite{cc1}. The trained channel charting maps are then fine-tuned using the labeled data with location information to acquire the absolute positions of the location of interest\cite{stephan2024angle}. Similar to the deep-learning-based positioning scheme, the positioning accuracy of channel charting is still highly related to the number of labeled data, the distributed of RPs, and the wireless environment variation within the target area.

\begin{figure*}[h]
\centering
\includegraphics[width=2\columnwidth]{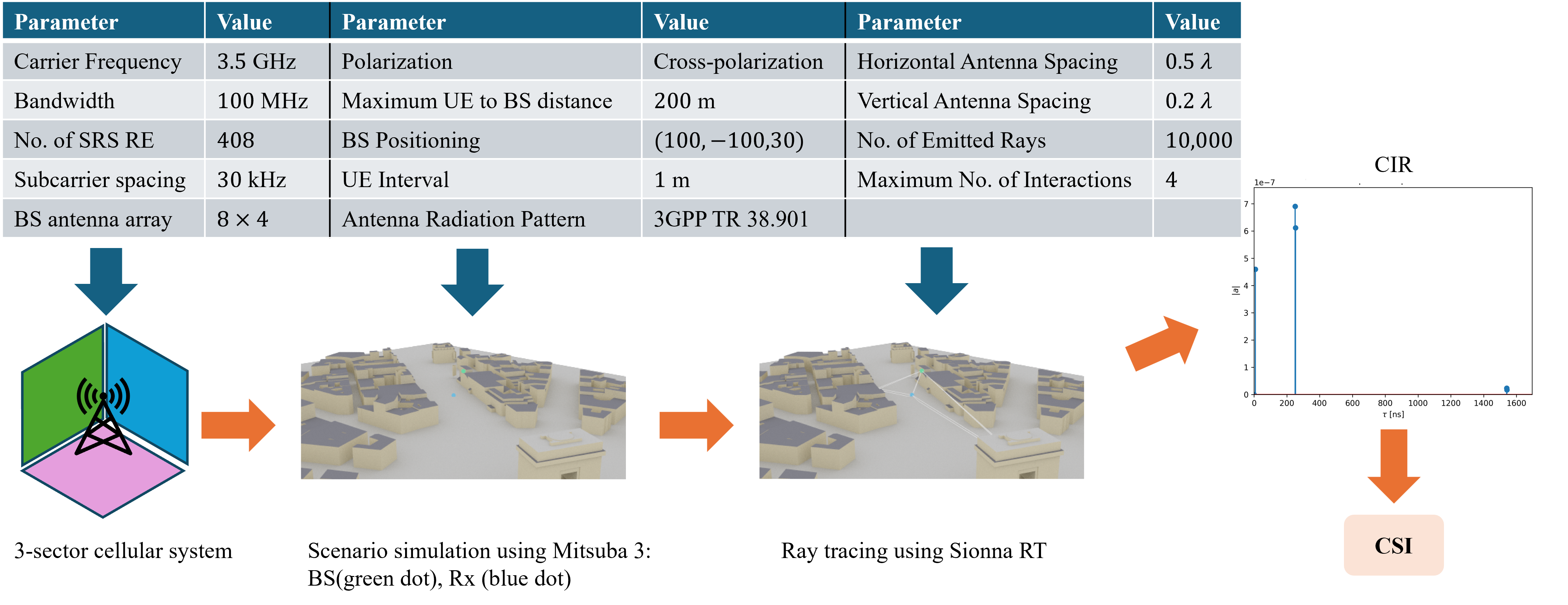}
\caption{Illustration of the 3-sector cellular system and the pipeline of CSI generation using Sionna RT and Mitsuba 3. Simulation settings are also demonstrated.}
\label{scenario}
\end{figure*}

To alleviate the dependency on the labeled data, the emerging semi-supervised learning framework only requires a partially labeled data set and achieves labeling efficiency by utilizing unlabeled examples for self-supervised training\cite{tarvainen2017mean,sohn2020fixmatch}. Inspired by this, \cite{gong2023semisupervised} proposed a semi-supervised scheme with contrastive learning to further extract high-dimensional features from the CSI data for positioning. However, practical factors are not considered in model design, such as variation in the environment and uneven distribution of RPs. 

To address the imperfections in the practical cellular positioning systems, we proposed three deep learning-based algorithms, namely semi-supervised framework with consistency, ensemble learning-based algorithm, and decoupled mapping heads-based algorithm. The proposed three algorithms swept the top three rewards in the 2024 Wireless Communication Algorithm Elite Competition sponsored by Huawei Ltd. The primary contributions of this paper are outlined as follows:

\begin{itemize}
\item To tackle the practical imperfection of limited labeld-data, we proposed a semi-supervised framework with consistency that could generate high-quality pseudo position labels to enlarge the labeled dataset for model training. This significantly increases the effective dataset size and improves localization accuracy. The method enables the model to concentrate on stable and accurate pseudo-labels during training, thereby demonstrating exceptional performance in CSI-based wireless positioning tasks.

\item To improve the accuracy and robustness of wireless positioning in dynamic positioning environments, we proposed an ensemble learning-based algorithm, which amalgamates the outputs from diverse pre-trained models. This model not only effectively mitigates the bias and variance issues that may arise from any single model but also capitalizes on the complementary strengths of diverse models.

\item To combat the challenge of unevenly-spaced anchor, we proposed a decoupled mapping heads-based algorithm that could exploit the sectorization characteristics of the 3-sector communication systems. This process involves rotating coordinates from other sectors by a specified angle to enhance data representation across all sectors. Three decoupled mapping heads are trained and dedicated for the positioning of each sector.  

\item Simulation results demonstrate that the proposed semi-supervised framework with consistency outperforms existing semi-supervised learning models widely used in the computation vision. The proposed three models outperform the benchmark position algorithms dramatically in terms of positioning error and robustness and the ensemble learning-based algorithm achieves the highest positioning accuracy at over half of the area of interest. 
\end{itemize}

\section{Practical imperfections and dataset for fingerprint-based positioning}

\subsection{Practical imperfections of fingerprint-based positioning}
Fingerprint-based positioning is a technique designed to determine a user's location by leveraging a pre-collected database of wireless signal fingerprints, such as Received Signal Strength (RSS) or Channel State Information (CSI). This method establishes a direct mapping between real-time signal measurements and spatial coordinates, offering impressive accuracy when the fingerprint database is both comprehensive and well-maintained. However, its practical deployment is hindered by the following significant practical imperfections that affect its scalability, flexibility, and resource efficiency.

\begin{itemize}
\item \textbf{limited-labeled dataset}: One key limitation is the scarcity of labeled data. Creating a fingerprint database requires meticulous signal measurements at known locations, a process that is labor-intensive and time-consuming. In most real-world scenarios, such as large indoor spaces like shopping malls or office complexes, it is impractical to collect data at every possible point. As a result, the number of localization-signal-mapping pairs remains small relative to the area being localized. This sparsity forces the system to rely on interpolation or estimation for unmeasured locations, often compromising accuracy, particularly in regions with few nearby reference points.

\item \textbf{Uneven-spaced anchors}: Another challenge arises from the uneven distribution of anchor points, which are the locations with known signal characteristics. Ideally, these points would be uniformly spaced to ensure consistent localization performance across the environment. In practice, however, logistical or environmental constraints, such as accessibility or signal propagation characteristics, often lead to clustered or sparse distributions. For example, open areas like hallways might have numerous anchor points, while confined spaces like small offices or corners have few. This unevenness results in variable accuracy, with well-sampled regions benefiting from precise localization and underserved areas suffering from degraded performance.

\item \textbf{Dynamic positioning environment}: The dynamic nature of wireless environments further complicates fingerprint-based positioning. Signal characteristics are highly sensitive to changes caused by factors such as multi-path propagation, line-of-sight (LoS) or non-line-of-sight (NLoS) conditions, and temporal variations. Multi-path effects occur when signals reflect off walls, furniture, or other objects, creating complex interference patterns that differ across locations and over time. Similarly, obstructions or the movement of people can shift a location from LoS to NLoS conditions, altering the measured signal profile. These environmental dynamics mean that a fingerprint database, once collected, may not remain valid for long. For instance, a database built in an empty office might fail to account for signal changes when employees arrive or furniture is rearranged. To counteract this, frequent updates to the database are necessary, but such maintenance demands significant resources, making the approach less viable for large-scale or rapidly changing settings.
\end{itemize}

\subsection{Dataset acquisition}
\label{data_generation}
To simulate conditions closely resembling practical cellular positioning scenarios, the wireless communication algorithm elite competition sponsored by Huawei Ltd. featured an outdoor fingerprint-based positioning that encompasses various scenarios under the above practical imperfections, i.e., limited-labeled dataset, uneven-space anchors, and dynamic positioning environment. As the dataset of the competition is not publicly available, we utilized Sionna RT\footnote{https://github.com/NVlabs/sionna-rt} and Mitsuba 3\footnote{https://www.mitsuba-renderer.org/} to generate a dataset with similar features to that used in the competition. Sionna RT provides accurate physical-layer modeling for wireless channel simulations through differentiable ray tracing, accounting for phenomena such as reflection, diffraction, and scattering. Mitsuba 3 handles scene rendering and processing, calculating intersections between rays and basic scene elements, such as triangles that form surface meshes.

Specifically, as illustrated in Fig. \ref{scenario}, we carried out simulation that replicates the single-site, three-sector configuration of the competition, with all parameters outlined on the top of Fig. \ref{scenario}. The BS located at (100, -100, 30) is equipped with $8 \times 4$ antenna array, while the Rxs with $2$ antennas are distributed across the three sectors at a height of $1.5$ m with a distance of $1$ m. Both the BS and Rxs adopt cross-polarization, and their antenna radiation pattern follows the 3rd generation partnership project (3GPP) TR 38.901. The carrier frequency is established at 3.5 GHz with a subcarrier spacing of 30 kHz. For a bandwidth of 100 MHz, adopting an 8-comb structure sounding reference signal (SRS) results in a total number of 408 SRS.

In Sionna RT, the scene comprises multiple objects that define the geometry and material characteristics and we adopted the 'etoile' scene for this simulation. As this scene is comprised of buildings with various sizes and heights, the environment of Rxs at different positions is distinct. This achieves the practical imperfection of dynamic positioning environment. The synthetic array option is enabled, which treats the array as a single entity during ray tracing, represented by a single antenna located at the center of the array. Phase shifts based on relative antenna positions are applied under a plane-wave assumption during channel impulse response (CIR) calculations to enhance simulation speed. We employed the Fibonacci method, configured to emit 10,000 rays from the transmitter path calculation. The ray directions are defined by Fibonacci grid points on a sphere, continuing their propagation until reaching the maximum number of reflections. In this simulation, we consider only direct and reflected paths.

For each Rx, the generated CIR data is further transformed into the frequency domain using a discrete Fourier transform and down-sampled by a factor of 4 in the subcarrier dimension. The resulting complex CSI matrix has dimensions of ($2 \times 32 \times 408$) (i.e., UE antennas $\times$ BS antennas $\times$ No. of SRSs). Each sample also includes two-dimensional (2D) coordinates and LOS/NLOS labels. The count of Rxs per sector is 21,745, 6,764, and 12,308, respectively. To mimic the practical imperfections of limited-labeled datasets and uneven-distributed anchors, we randomly select 10\% of Rxs with two-dimensional (2D) coordinates.

To more accurately simulate the process of obtaining CSI in a real environment, this paper considers the effects of time advance (TA) and additive white Gaussian noise (AWGN) on CSI. Specifically, TA results in an overall shift of the power delay profile (PDP) curve in the time domain, which manifests as a phase offset in the frequency domain. The inclusion of AWGN is achieved by adding noise to each CSI, which has the same dimensions as the original CSI and follows a complex Gaussian distribution.

\begin{figure*}[h] 
 \centering 
 \includegraphics[width=2\columnwidth]{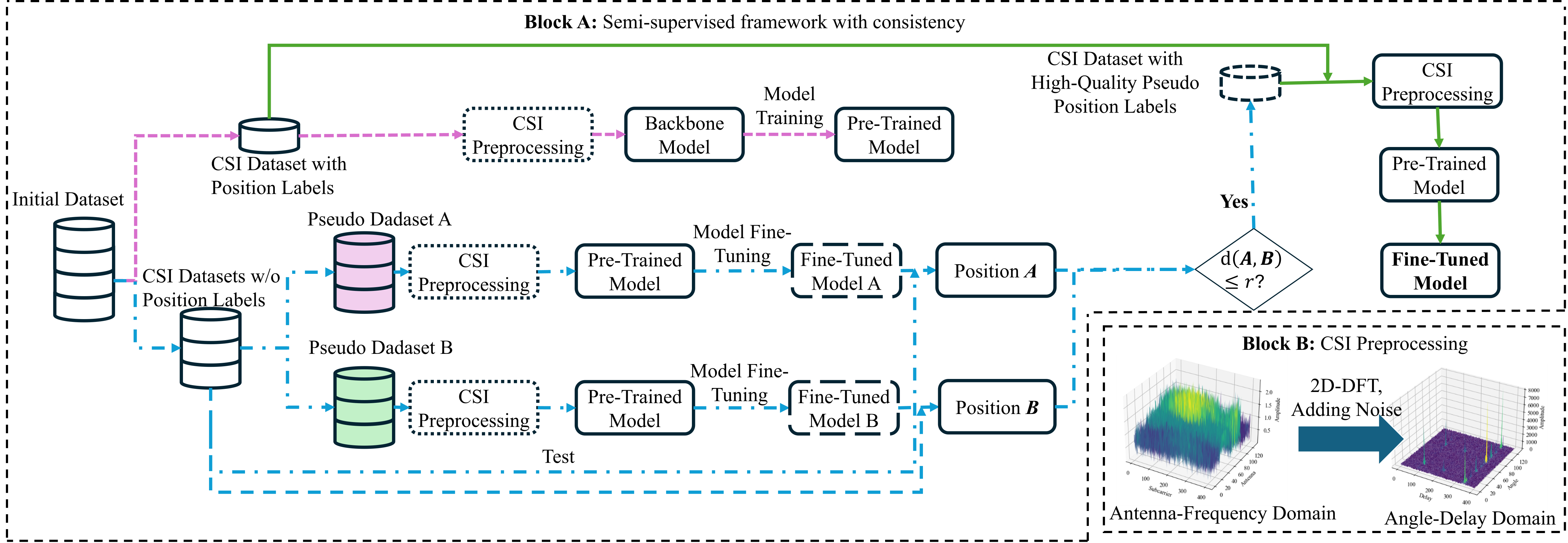} 
 \caption{Illustration of the semi-supervised framework with consistency (Block A) and CSI preprocessing (Block B). The main idea of the semi-supervised framework with consistency is generating high-quality pseudo position labels to enlarge the labeled dataset. 
 \textit{Stage 1} (denoted as dashed line): Using labeled dataset to obtain the pre-trained model; 
 \textit{Stage 2} (denoted as dot-dashed line): Using different interpolated methods to generate two pseudo datasets, then fine-tune the pre-trained models. Predicting the position of the CSI data without position labels. If the predicted positions of the same CSI is smaller than a threshold, the high-quality pseudo position label is denoted as the mean of the predicted positions. 
 \textit{Stage 3} (denoted as solid line): Combining the labeled dataset and CSI data with high-quality pseudo labels to fine-tune the pre-trained model and obtain the final model for position. In CSI preprocessing, CSI is transformed from the antenna-frequency domain into the angle-delay domain and added with noise.} 
\label{Fig:semi-supervised}
\end{figure*}

\section{Key enabling techniques} 
To improve positioning performance, we proposed three deep-learning-based schemes to resolve the above practical imperfections, i.e., limited-labeled dataset, uneven-space anchors, and dynamic positioning environment, respectively.

\subsection{Semi-supervised framework with consistency} 

To tackle the practical imperfection of limited labeld-data, we proposed a semi-supervised framework with consistency that could generate high-quality pseudo position labels to enlarge the labeled dataset for model training. As illustrated in Fig. \ref{Fig:semi-supervised}, the semi-supervised framework with consistency comprises of 3 stages.

\textit{Stage 1}: A pre-trained model is obtained by training the backbone model using the CSI data with position labels. Notably, prior to training, the CSI is preprocessed by transforming from the antenna-frequency domain into the angle-delay domain using 2D discrete Fourier transform (2D-DFT) and adding additional noise. The CSI in the angle-delay domain exhibits notable sparsity, which is advantageous for model training. 

\textit{Stage 2}: High-quality pseudo position labels are generated based on the idea of consistency. Specifically, two pseudo datasets A and B are generated using different interpolation methods based on the CSI data without position labels. The pre-trained model obtained in \textit{Stage 1} is fine-tuned using pseudo datasets A and B, and leads to fine-tuned models A and B, respectively. The positions of the CSI data without position labels are predicted using the fine-tuned model A and B and denoted as position $\textbf{A}$ and $\textbf{B}$. If the positions of the same CSI data in position $\textbf{A}$ and $\textbf{B}$ are closer than a pre-defined Euclidean distance $r$, we denote the high-quality pseudo position label of this CSI as the mean of two predicted positions. In this way, we generate a CSI dataset with high-quality pseudo position labels. The core idea of this design is inspired by the predictive consistency\cite{sohn2020fixmatch}. As the training dataset (i.e., pseudo dataset A and B) used to fine-tune the pre-trained model originated from the same dataset with minor variations, their predictions for the same test set should converge toward the ground-truth values or exhibit close proximity, i.e., predictive consistency. 

\textit{Stage 3}: We further fine-tune the pre-trained model using the combination of the CSI dataset with position labels and  with high-quality pseudo position labels obtained in \textit{Stage 2}. The fine-tuned model is finally used for position tasks.

The performance of the semi-supervised framework with consistency could be further enhanced by performing \textit{Stage 2} and \textit{Stage 3} iteratively to generate a larger volume of CSI dataset with high-quality pseudo position labels. 

\begin{figure*}[h]
\centering
\includegraphics[width=2\columnwidth]{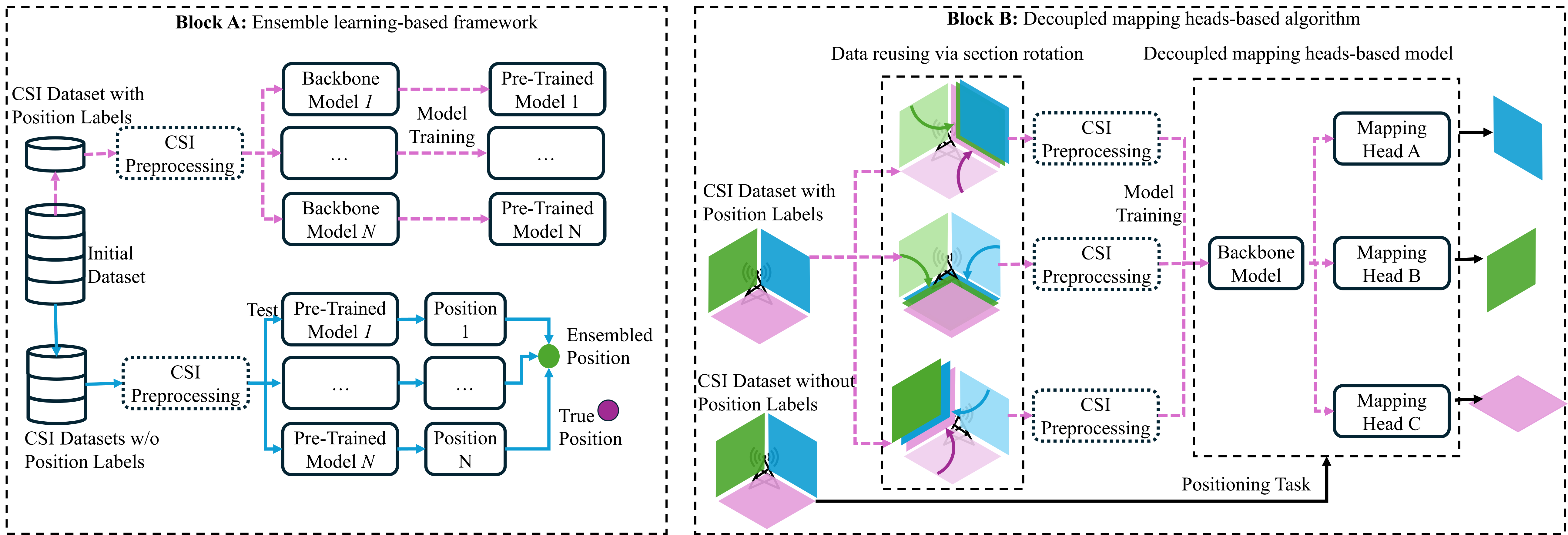}
\caption{Illustration of the ensemble learning-based algorithm (Block A) and decoupled mapping heads-based algorithm (Block B). The ensemble learning-based algorithm amalgamates the outputs from diverse pre-trained models, which not only effectively mitigates the bias and variance issues that may arise from any single model but also capitalizes on the complementary strengths of diverse models. The decoupled mapping heads-based algorithm exploits the sectorization characteristics of the 3-sector communication systems by rotating coordinates from other sectors by a specified angle to enhance data representation across all sectors. Three decoupled mapping heads are trained and dedicated for the position of each sector.  } 
\label{Fig:ensemble_decouple}
\end{figure*}

\subsection{Ensemble learning-based framework}

To improve the accuracy and robustness of wireless positioning in dynamic positioning environments, we proposed an ensemble learning-based algorithm, which comprises of 2 stages illustrated in the \textbf{Block A} of Fig. \ref{Fig:ensemble_decouple}

\textit{Stage 1}: Prior to training, the CSI data with position labels are preprocessed the same way as before. Then, diverse backbone models are trained using the preprocessed dataset. Variations are introduced by adjusting the model parameters, architectures, and hyperparameters. Although these models are trained on the same training dataset, differences in their internal structures and settings lead to subtle variations in their responses to the test datasets.

\textit{Stage 2}: In the ensemble learning stage, the final positions are generated by amalgamating the outputs from diverse pre-trained models, such as weighted averaging or a voting mechanism. This model not only effectively mitigates the bias and variance issues that may arise from any single model but also capitalizes on the complementary strengths of diverse models. The ensemble learning stage leverages consensus regions among the models. These are the data points for which multiple models yield similar predictions, thus constructing a more stable and accurate final prediction. Consequently, it enhances the overall positioning accuracy and robustness in dynamic environments.

\subsection{Decoupled mapping heads-based algorithm}

To combat the challenge of unevenly-spaced anchors, we proposed to exploit the sectorization characteristics of the 3-sector communication systems. This process involves rotating coordinates from other sectors by a specified angle to enhance data representation across all sectors. As illustrated in the \textbf{Block B} of Fig. \ref{Fig:ensemble_decouple}, we proposed a decoupled mapping heads-based model, which employs a coupled training strategy alongside decoupled inference, effectively leveraging the idea of sector complementation. The main stages of the decoupled mapping heads-based model are sector rotation and decoupled mapping heads.

\textbf{Sector rotation}: 
To tackle the uneven-spaced anchor challenge, we propose to perform sector rotation. By rotating the sector, anchors originally concentrated in a specific area can cover sparse regions of the current sector, achieving a more uniform distribution. Only the direction coordinates of the anchors are altered, while key physical characteristics like signal strength and time delay, related to distance, remain unchanged. Since the actual propagation distance between anchors and user equipment does not change, the signal features still accurately reflect the target location. This ensures that the rotated data augmentation does not introduce physical contradictions, thus offering the model more reliable data support. By rotating unlabeled data from adjacent sectors (e.g., rotating dataset in sector B to supplement dataset in sector A), we construct a virtual sector dataset tripling the original size.

\textbf{Decoupled mapping heads}: The three decoupled mapping heads are basically fully connected (FC) layers. Recognizing the three-sector distribution of cellular BSs, this approach introduces three decoupled mapping heads, allowing for independent position mapping in each sector. This design retains the unique characteristics influenced by environmental differences while also promoting shared feature extraction to reduce the overall number of network parameters.

\section{Experimental evaluation}

\begin{figure*} \centering 
\subfigure[Cumulative probability of error distance] { 
\label{Error_distance_cdf}     
\includegraphics[width=0.98\columnwidth]{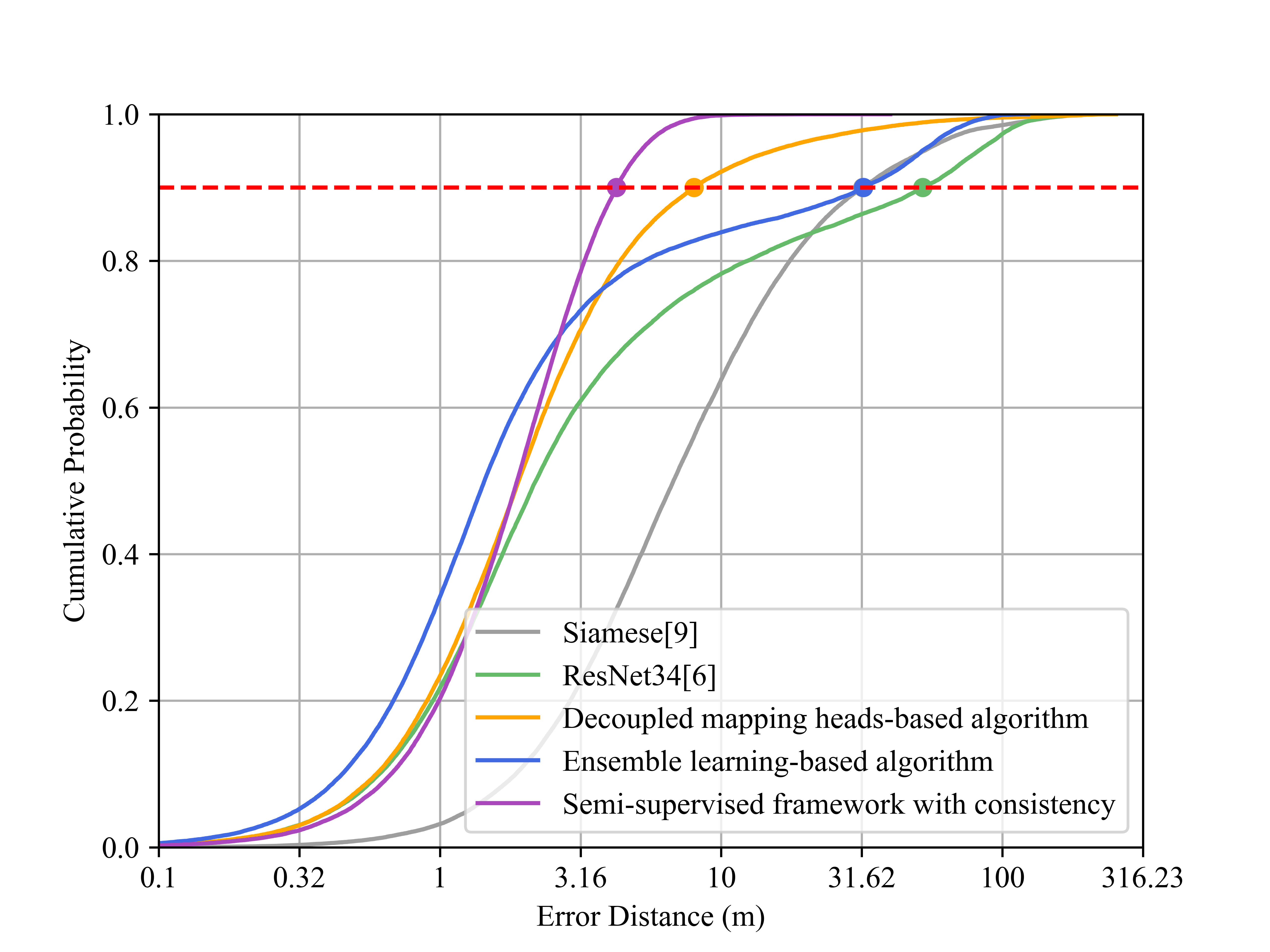} 
}
\subfigure[Error distance at various percentile] { 
\label{Error_distance_probability}     
\includegraphics[width=0.98\columnwidth]{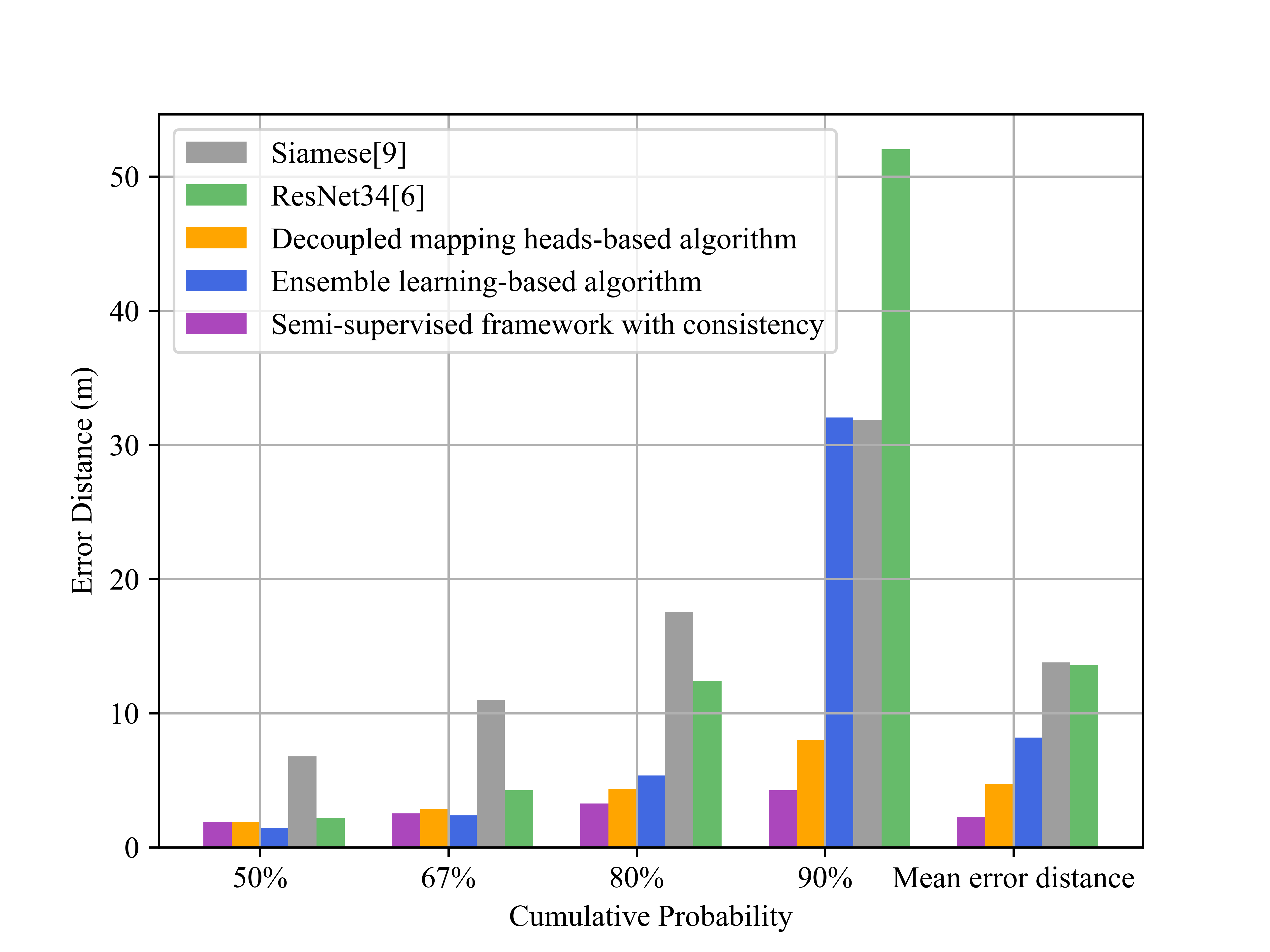} 
} 
\caption{Comparison of various localization algorithms
}   
\label{data}     
\end{figure*} 

\subsection{Experimental Setup}
This study evaluates the proposed algorithms using the generated dataset described in Section \ref{data_generation}. The positioning error is defined as the Euclidean Distance between predicted and true coordinates: $d_i = \sqrt{(\hat{x}_i - x_i)^2 + (\hat{y}_i - y_i)^2}$ where $\hat{x}_i, \hat{y}_i$ are predicted coordinates and $x_i, y_i$ are ground-truth coordinates. 
\begin{table}[t]
\centering
\caption{Comparison of backbone models}
\label{tab:backbone}
\begin{tabular}{cccc}
\hline
Model & Parameters (M) & FLOPs (G) & Distance Error (m) \\
\hline
ShuffleNetV2\cite{ma2018shufflenet} & 1.26 & 0.08 & 16.80 \\
MobileNetV2\cite{sandler2018mobilenetv2} & 2.23 & 0.17 & 14.61 \\
ResNet18 & 11.18 & 0.98 & 14.25 \\
ResNet34 & 21.29 & 1.95 & 13.36 \\
ResNet50 & 23.52 & 2.19 & 13.23 \\
ResNet101 & 42.51 & 4.17 & 12.63 \\
\hline
\end{tabular}
\end{table}
\subsection{Backbone Selection}
In the realm of computer vision (CV), the ResNet family, widely recognized as the most common backbone network, leverages residual learning to effectively tackle the gradient vanishing problem and significantly boost feature extraction capabilities. As demonstrated in Table \ref{tab:backbone}, ResNet34\cite{he2016deep} exhibits a remarkable 8.5\% lower error rate compared to MobileNetV2\cite{sandler2018mobilenetv2}. Despite utilizing fewer parameters and computational resources, it incurs only a 6\% performance cost yet surpasses both ResNet50 and ResNet101. The advantageous balance ResNet34 strikes between efficiency and accuracy renders it a superior choice among various backbone options. Consequently, to eliminate confounding factors associated with differences in backbone capacity, ResNet34 was consistently employed across all experiments.
\subsection{Comparison with Existing Semi-Supervised Methods}
We compare the mean error distance of the proposed semi-supervised framework with consistency and other typical semi-supervised learning approaches that are widely used in the CV domain. In Table. \ref{tab:semi}. It is observed that the proposed semi-supervised framework with consistency outperforms Mean Teacher \cite{tarvainen2017mean} and FixMatch \cite{sohn2020fixmatch} dramatically. 

\textbf{Superiority over Mean Teacher}: Mean teacher is a typical CV model and its superior performance relies heavily on CV-specific augmentations (e.g., random cropping/flipping for images), which are inapplicable to CSI. As a result, the mean error distance of Mean Teacher is significantly larger than that of the proposed showed a mean error distance of 7.55 m due to its inherent dependence on strong augmentation effects. 

\textbf{Superiority over FixMatch}: FixMatch's original confidence thresholding (using class probability $> 0.95$) proves ineffective for our regression task as shown by its 7.28 error because we do not have the class information. We innovatively redesigned the filtering criterion using coordinate prediction errors, establishing a distance-based threshold (5.65 vs 7.28 error). This fundamental paradigm shift from classification confidence to regression accuracy constitutes a principal advancement for positioning applications.
\begin{table}[h]
\centering
\caption{A comparison between the proposed semi-supervised framework with consistency (denoted as "ours" in the table) and other typical semi-supervised learning approaches that are widely used in the CV domain.}
\label{tab:semi}
\begin{tabular}{cc}
\hline
Method & Mean Error Distance  (m) \\
\hline
Mean Teacher \cite{tarvainen2017mean}& 7.55 \\
FixMatch \cite{sohn2020fixmatch}& 7.28 \\
FixMatch (distance threshold) & 5.65 \\
Ours & 4.13 \\
Ours (iterative) & 2.23 \\
\hline
\end{tabular}
\end{table}

\textbf{Innovations of the proposed method}: Unlike conventional semi-supervised learning paradigms, our framework introduces a novel model coordination mechanism that systematically implements cross-consistency verification through dual-network collaboration. The architecture employs two structurally divergent networks that independently generate pseudo-labels during the post-training phase, with threshold filtering. This iterative refinement process enables progressive optimization of the learning system, yielding significant performance improvements as evidenced by a 42.6\% error reduction (from 4.13 m to 2.23 m).

\subsection{Comparison with Existing Positioning Methods}
Fig. \ref{Error_distance_cdf} illustrates the CDF of the error distance of various positioning algorithms qualitatively with the following observations. The overall observation is that the proposed semi-supervised consistency algorithm and the ensemble learning-based algorithm outperform the rest algorithms dramatically. Particularly, it is observed that the proposed semi-supervised consistency algorithm is the only one to achieve sub-\SI{10}{\metre}, which outperforms its counterparts, of which the worst case positioning errors reach up to over \SI{100}{\metre}. This indicates that the proposed semi-supervised consistency algorithm provides a relatively-stable positioning performance over the area of interest. Additionally, the ensemble learning-based algorithm can achieve the lowest positioning error in around 70\% of the area of interest, while the positioning error worse off dramatically in the rest 30\% of the area of interest. This indicates that the robustness of the ensemble learning-based positioning algorithm is its main disadvantage.

To further quantitatively compare the positioning performance, compliant with the 3GPP TR 38.843 \cite{38_843}, we adopted [90\%, 80\%, 67\%, 50\%] CDF percentiles to analyze the performance of positioning, which is illustrated in Fig. \ref{Error_distance_probability}. Consistent with the observation and analysis of the semi-supervised consistency algorithm in Fig. \ref{Error_distance_cdf}, the semi-supervised consistency algorithm achieves the most stable positioning at the percentile of interest and the lowest positioning error at 90\% percentile. The positioning errors of the rest 4 algorithms grow dramatically with the increase of percentile, which indicates the poor robustness of these algorithms. 

\section{Future research direction}
\subsection{Computational Complexity and Inference Speed}
Current research on deep learning-based cellular positioning emphasizes primarily positioning accuracy while often overlooking computational complexity and inference speed during model design and performance evaluation. This oversight limits the practical applicability of existing algorithms. The 3GPP TR 38.843 \cite{38_843} underscores the urgent need to consider the computational capacity and inference speed of these algorithms. The perspective approaches to reduce the computational complexity of deep learning models are summarized as follows:

\textbf{Model Pruning}: The redundant or less significant weights, layers, or neurons from the model could be removed to decrease the number of parameters, thereby reducing its computational complexity. Model pruning could be performed in 
\begin{inparaenum}[\itshape a\upshape)]
\item weight level: eliminating weights below a threshold based on magnitude;
\item neuron level: removing neurons with low activation or contribution to the output;
\item layer level: simplifying architectures by removing redundant layers.
\end{inparaenum}

\textbf{Knowledge Distillation}: Knowledge distillation enables effective compression of complex teacher models into lightweight student networks while preserving model efficacy with substantially reduced computational costs.

\subsection{Generalization Ability}
Generalization performance is a crucial indicator of the effectiveness of AI models across various scenarios, highlighted by the emphasis on this aspect in 3GPP TR 38.843 \cite{38_843}. To further improve the generalization of fingerprint-based positioning models, the following perspective approaches can be considered.

\textbf{Multi-Task Learning}: As discussed in Section I, the signal propagation is very sensitive to the physical environment, including LoS/NLoS, outdoor urban/rural, the knowledge of the environment can be beneficial to improve the generalization of the positioning model. To achieve this, multi-task learning is a promising candidate to leverage inter-task (e.g., channel identification and positioning) correlations by optimizing the performance of multiple tasks. 

\textbf{Dynamic Data Augmentation}: Generating diverse simulated data in real-time during training (e.g., multipath effects, signal occlusion, environmental noise) to improve the generalization of the model to complex channel conditions and unknown scenarios. For instance, more real-world uncertainties could be simulated by randomly perturbing wireless signal features (e.g., delay, angle).

\section{Conclusion}
To tackle the practical imperfections of fingerprint-based positioning, i.e., limited labeled-dataset, dynamic wireless environments within the target, and unevenly-spaced anchors, we developed three innovative positioning frameworks, namely the semi-supervised framework with consistency, ensemble learning-based algorithm, and decoupled mapping heads-based algorithm. Specifically, the semi-supervised framework with consistency effectively generates high-quality pseudo-labels, enlarging the labeled-dataset for model training. The ensemble learning-based algorithm amalgamates the positioning coordinates from models trained under different strategies, effectively combating the dynamic positioning environments. The decoupled mapping heads-based algorithm utilized sector rotation scheme to resolve the uneven-spaced anchor issue. Comprehensive simulation results demonstrate the superior performance of our proposed positioning algorithms.

\bibliographystyle{ieeetr}
\bibliography{Myreference}

\end{document}